\documentclass[a4paper,twocolumn,prl,showpacs]{revtex4}
\usepackage{graphicx,amsmath,amssymb}

\renewcommand{\[}{\begin{equation}}
\renewcommand{\]}{\end{equation}}
\def\bea{\begin{eqnarray}}
\def\eea{\end{eqnarray}}
\def\nn{\nonumber\\}
\newcommand{\equ}[1]{Eq.~(\ref{#1})}
\newcommand{\eqs}[2]{Eqs.~(\ref{#1}) and (\ref{#2})}

\newcommand{\evo}{\langle \omega^2 \rangle}

\newcommand{\zz}{{\cal Z}}
\newcommand{\rr}{{\cal R}}

\def\runtime{(\the\time)\qquad\the\month/\the\day/\the\year}
\def\today
 {\count10=\year\advance\count10 by -2000 \number\day--\ifcase
  \month \or Jan\or Feb\or Mar\or Apr\or May\or Jun\or
             Jul\or Aug\or Sep\or Oct\or Nov\or Dec\fi--\number\count10}

\def\hour{\count10=\time\count11=\count10
\divide\count10 by 60 \count12=\count10
 \multiply\count12 by 60 \advance\count11 by -\count12\count12=0
\number\count10 :\ifnum\count11 < 10 \number\count12\fi\number\count11}

\begin{document}

\title{Lyddane-Sachs-Teller relationship in linear magnetoelectrics}

\author{Raffaele Resta}

\affiliation{Dipartimento di Fisica, Universit\`a di Trieste, Italy, \\
and DEMOCRITOS National Simulation Center, IOM-CNR, Trieste, Italy}

\date{}

\begin{abstract}
In a linear magnetoelectric the lattice is coupled to electric and magnetic fields: both  affect the longitudinal-transverse splitting of zone-center optical phonons on equal footing. A  response matrix relates the macroscopic fields $(D,B)$  to  $(E,H)$ at infrared frequencies. 
It is shown that the response matrices at frequencies $0$ and $\infty$ 
fulfill a generalized Lyddane-Sachs-Teller relationship. The rhs member of such relationship is expressed in terms of weighted averages over the longitudinal and transverse excitations of the medium, and assumes a simple form for an harmonic crystal.
\end{abstract}


\maketitle \bigskip\bigskip

The original Lyddane-Sachs-Teller (LST) relationship~\cite{Lyddane41} applies to the simple case of  
a cubic binary crystal in the harmonic regime. It relates four macroscopically measurable constants as \[ \frac{\varepsilon(0)}{\varepsilon(\infty)} = \frac{\omega^2_{\rm L}}{\omega^2_{\rm T}} . \label{lst} \] Here $\varepsilon(0)$ is the static dielectric constant, which includes the lattice contribution, $\varepsilon(\infty)$ is the so called ``static high frequency'' (a.k.a ``clamped-ion'') dielectric constant, which accounts for the electronic response only, and $\omega_{\rm L}$ ($\omega_{\rm T}$) is the zone-center  longitudinal (transverse) optical frequency \cite{textbook}. It is remarkable that all {\it microscopic} parameters (force constants, masses, Born effective charges, cell volume) disappear from \equ{lst}. 

Magnetoelectrics (MEs) are insulators where electric fields control magnetization, and conversely magnetic fields control polarization; they attracted considerable theoretical and technological interest in recent times~\cite{Fiebig05,Eerenstein06,Iniguez08,Hehl08,Hehl09,Essin09,Wojdel09,Essin10,Wojdel10}. The simplest and most studied single-crystal linear ME is antiferromagnetic Cr$_2$O$_3$ \cite{Fiebig05,Iniguez08,Hehl08,Hehl09}.
In any linear ME the role of the dielectric function $\varepsilon(\omega)$ is played by the $2\times2$ response matrix---called $\rr(\omega)$ here---which yields the macroscopic fields $(D,B)$ in terms of $(E,H)$ at frequency $\omega$: \[ \left(\begin{array}{c} D \\ B \end{array} \right) = \rr(\omega) \left(\begin{array}{c} E \\ H \end{array} \right) \equiv  \left(\begin{array}{cc} \varepsilon(\omega) & \alpha(\omega) \\  \alpha(\omega)  & \mu(\omega) \end{array} \right)  \left(\begin{array}{c} E \\ H \end{array} \right) , \label{resp} \] where $\varepsilon$, $\mu$, and $\alpha$ are  permittivity, magnetic permeability, and ME coupling, respectively.
In this Letter we are going to show that a generalized LST relationship holds in the form \[  \frac{\mbox{tr } \{ \rr^{-1}(\infty) \rr(0) \} - 2}{2 - \mbox{tr } \{ \rr^{-1}(0) \rr(\infty) \} }  =  \frac{\omega^2_{\rm L}}{\omega^2_{\rm T}}  . \label{rap} \] 
In the simple case of a magnetically inert material (i.e. $\mu \equiv \mbox{constant}$, $\alpha \equiv 0$) the lhs of \equ{rap} equals indeed $\varepsilon(0)/\varepsilon(\infty)$. While in ordinary dielectrics the LO-TO splitting is due to the coupling of the ionic displacements to macroscopic electric fields, in linear MEs it is due to the coupling of both (electric and magnetic) fields on the same footing: this is perspicuous in the lhs of \equ{rap}.

The simple form of \eqs{lst}{rap} requires a crystalline system with only a single IR-active mode at the zone center.  The additional requirement of cubic symmetry can be relaxed; it is nonetheless convenient to consider only crystals whose symmetry is orthorombic or higher; then all crystalline  tensors can be simultaneously diagonalized (as e.g. in  Cr$_2$O$_3$). This allows us to  adopt simple scalar-like notations, as in \eqs{lst}{rap} with the proviso that we deal separately with each principal direction. 

Over the years the LST relationship has been extended in several ways, to cover cases where more than one  IR-active mode per direction exists~\cite{Kurosawa61,Cochran62}, the crystal is in a low-symmetry class~\cite{Lax71,Gonze97}, or even the material is noncrystalline and/or anharmonic~\cite{Barker75,Barker75b,Noh89,Sievers90}. The general case can be written as  \[ \frac{\varepsilon(0)}{\varepsilon(\infty)} = \frac{\evo_{\rm L}}{\evo_{\rm T}} . \label{lst2} \] The quantities in the rhs are weighted averages, obtained from moments of the appropriate spectral functions. The derivation is based on general principles of statistical mechanics and does not require an Hamiltonian, even less an harmonic one~\cite{Barker75,Noh89}; in the special case of a single IR-active harmonic mode per direction \equ{lst2} is equivalent to \equ{lst}. In this Letter  we generalize the viewpoint of Ref.~\cite{Noh89} to the ME case, showing that \[  \frac{\mbox{tr } \{ \rr^{-1}(\infty) \rr(0) \} - 2}{2 - \mbox{tr } \{ \rr^{-1}(0) \rr(\infty) \} }  = \frac{\evo_{\rm L}}{\evo_{\rm T}} ,\label{lst3} \] where the rhs is defined below, \eqs{m1}{m2}.

The presentation proceeds as follows. We start at a very general level without any {\it microscopic} assumption about the ME medium, and using only very general principles in order to arrive at \equ{lst3}. We will then apply the general results to a crystalline system in the harmonic regime, and finally we will show that \equ{lst3} reduces to \equ{rap} in the single-mode case.

We write explicitly the linear response matrix of the ME medium as the sum of its real and imaginary part: $\rr(\omega) = \rr'(\omega) +i \rr''(\omega)$, and analogously for its inverse; both $\rr(\omega)$ and $\rr^{-1}(\omega)$ obey the Kramers-Kronig relationships in the form \bea \rr'(\omega) - \rr(\infty) &=& \frac{2}{\pi} \int_0^\infty d \omega' \; \frac{\omega' \rr''(\omega')}{\omega'^2 - \omega^2} , \label{k1} \\ {\rr^{-1}}'(\omega) - \rr^{-1}(\infty) &=& \frac{2}{\pi} \int_0^\infty d \omega' \; \frac{\omega' {\rr^{-1}}''(\omega')}{\omega'^2 - \omega^2} . \label{k2} \eea From these, it follows immediately that the numerator and denominator in the lhs of \eqs{rap}{lst3} are
\[ \mbox{tr} \{ \rr^{-1}(\infty) \rr(0) \} - 2 =  \frac{2}{\pi} \int_0^\infty  \frac{d\omega}{\omega} \mbox{tr} \{ \rr^{-1}(\infty) \rr''(\omega) \} , \label{kk1}\] 
\[ 2 - \mbox{tr} \{ \rr^{-1}(0) \rr(\infty) \}  =   - \frac{2}{\pi} \int_0^\infty \frac{d\omega }{\omega} \mbox{tr} \{ {\rr^{-1}}''(\omega) \rr(\infty) \} . \label{kk2}\] The integrands in the rhs of \eqs{kk1}{kk2} are interpreted here as the transverse and longitudinal spectral weights, respectively, by means of which we {\it define} the second moments \bea
\evo_{\rm T} &=& \frac{\int_0^\infty  \frac{d\omega }{\omega} \omega^2 \mbox{tr } \{ \rr^{-1}(\infty) \rr''(\omega) \} }{ \int_0^\infty \frac{d\omega }{\omega} \mbox{tr } \{ \rr^{-1}(\infty) \rr''(\omega) \} } \label{m1} \\ \evo_{\rm L} &=&  \frac{\int_0^\infty \frac{d\omega }{\omega} \omega^2 \mbox{tr } \{ {\rr^{-1}}''(\omega) \rr(\infty) \} }{ \int_0^\infty \frac{d\omega }{\omega} \mbox{tr } \{ {\rr^{-1}}''(\omega) \rr(\infty) \} } \label{m2} . \eea The reason for the semantics ({\it transverse} and {\it longitudinal}) may appear obscure at this point; it will become clear when specializing \eqs{m1}{m2} to an harmonic crystal---see also \equ{mean} below.

In order to arrive at our main result, \equ{lst3}, we exploit the ``superconvergence'' theorem~\cite{Altarelli72}. For large $\omega$ (i.e. for $\omega$ much larger than all the resonances of the medium) $\rr''(\omega)$ vanishes; \eqs{k1}{k2} yield, to leading order in $1/\omega^2$,
 \[ \rr(\infty)^{-1}  \rr(\omega)   \simeq {\cal I} - \frac{2}{\pi \omega^2} \int_0^\infty d \omega' \; \omega' \rr(\infty)^{-1}  \rr''(\omega') , \label{kkk1}  \]   \[ \rr(\omega)^{-1}  \rr(\infty)   \simeq {\cal I} - \frac{2}{\pi \omega^2} \int_0^\infty d \omega' \; \omega'  {\rr^{-1}}''(\omega') \rr(\infty), \label{kkk2}  \]  where ${\cal I}$ is the $2\times 2$ identity. Inversion of \equ{kkk1} to the same order gives the alternative expression
\[ \rr(\omega)^{-1}  \rr(\infty)   \simeq {\cal I} + \frac{2}{\pi \omega^2} \int_0^\infty d \omega' \; \omega' \rr(\infty)^{-1}  \rr''(\omega') . \label{inv}  \] Next we take the trace of \eqs{kkk2}{inv}; permuting the matrices in the product we get the identity \bea & & \int_0^\infty d \omega \; \omega \; \mbox{tr } \{ \rr(\infty)^{-1}  \rr''(\omega)\} \nn &=& -
\int_0^\infty d \omega \; \omega \; \mbox{tr } \{ {\rr(\omega)^{-1}}''  \rr(\infty)\} . \label{iden} \eea In order to arrive at  \equ{lst3} it is enough to put together Eqs.~(\ref{kk1}), (\ref{kk2}), (\ref{m1}), (\ref{m2}), and (\ref{iden}).

We stress that at the root of the superconvergence identity, \equ{iden}, is the assumption that $\omega = \infty$ actually means $\omega$ much higher than all the frequencies of ionic motions, yet lower than the frequencies of electronic excitations \cite{textbook}. Therefore the clamped-ion response $\rr(\infty)$ is a real matrix.

Next we address an harmonic  crystal. A zone-center optical mode is lattice-periodical;  it is then expedient to consider the energy per cell of a (macroscopically homogeneous) solid with a frozen-in phonon distortion.  This energy is well defined only when a prescription for taking the thermodynamic limit is given. We cut a sample in the shape of a slab parallel to the principal axes; we remind our assumption that all crystal tensors are diagonal on them. If the slab is  free-standing in vacuo,  all fields vanish outside ($E$=$D$=$H$=$B$=$0$), while the value of the fields inside depend on the polarization of the mode. Simple electrostatics and magnetostatics imply that if the phonon polarization is parallel to the slab (``transverse''), both $E$ and $H$ vanish, while if it is perpendicular (``longitudinal'')  $D$ and $B$ vanish \cite{rap_a30}. The order of the limits (first a slab, then its thickness going to infinity) is essential, and the two energies (longitudinal and transverse) are indeed different in the thermodynamic limit. Similar reasonings apply if the lattice-periodical mode is regarded as  the ${\bf k} \rightarrow 0$  limit of a finite-${\bf k}$ optical phonon \cite{textbook}.

If the crystal has $N$ IR-active modes in each principal direction,
we denote with $\omega_n$ the zone-center TO frequencies (i.e those with $E$=0 and $H$=0); equivalently, $\omega_n^2$ are the eigenvalues of {\it the analytical part} of the dynamical matrix at ${\bf k} = 0$.
The free energy per cell in function of the normal coordinates $u_n$, $E$, and $H$ (taken as independent variables)  is expanded to second order as~\cite{Fiebig05,Iniguez08} \bea  && F(E,H,\{ u_n \}) = F_0 + \frac{1}{2} \sum_n \omega_n^2 u_n^2  \nn &-& \frac{\Omega}{8 \pi} [ \, \varepsilon(\infty) E^2 + 2 \alpha(\infty)  E H + \mu(\infty)  H^2 \, ]  \nn &-& \sum_n ( u_n Z_n^* E + u_n \zeta^*_n H ) , \label{free} \eea where we adopt atomic Gaussian units \cite{units}, and $\Omega$ is the cell volume. 
$Z_n^*$ are the mode-effective charges and $\zeta_n^*$ their magnetic analogues; notice that in ordinary units the normal-mode coordinates would include a factor with the dimensions of (mass)$^{1/2}$, while $Z_n^*$ and $\zeta_n^*$ would include a factor with the dimensions of (mass)$^{-1/2}$.

The derivatives of $F$ provide the equations of motion in the form \bea D &=& - \frac{4 \pi}{\Omega} \frac{\partial F}{\partial E} = \varepsilon(\infty) E +  \alpha(\infty) H + \frac{4 \pi }{\Omega} \sum_n Z_n^* u_n  \nn
B &=& - \frac{4 \pi}{\Omega} \frac{\partial F}{\partial H} = \alpha(\infty) E + \mu (\infty) H + \frac{4 \pi }{\Omega} \sum_n \zeta_n^* u_n
\nn f_n &=& - \frac{\partial F}{\partial u_n} = - \omega_n^2 u_n  + Z_n^* E + \zeta_n^* H. \label{motion} \eea 
We then consider forced oscillations at frequency $\omega$, i.e. $f_n = -\omega^2 u_n$. Elimination of the $u_n$'s from \equ{motion} provides the linear ME response, including the lattice contribution; we cast it in compact form as \bea \rr'(\omega) &=& \rr(\infty) +  \frac{4 \pi}{\Omega} \sum_n  \frac{\zz_n \zz_n^\dagger}{\omega_n^2  - \omega^2}  \label{ini} \\ \rr''(\omega) &=&  \frac{4 \pi^2}{\Omega} \sum_n  \zz_n \zz_n^\dagger \, \delta(\omega_n^2 - \omega^2) , \eea 
where the ME lattice coupling vectors are
\[  \zz_n = \left(\begin{array}{c}  Z^*_n \\ \zeta^*_n \end{array} \right) , \qquad
\zz_n^\dagger = \left(\begin{array}{cc}  Z^*_n & \zeta^*_n \end{array} \right) .
\]  \equ{ini} is the elegant result obtained in 2008 by J. \`I\~niguez~\cite{Iniguez08}; the TO frequencies $\omega_n$ are clearly the $N$ poles of $\rr(\omega)$. 

Actual computations performed for the paradigmatic material Cr$_2$O$_3$ and based on \equ{ini}  show that the magnetoelectric coupling $\alpha(0)$ is significantly enhanced by the lattice contribution \cite{Iniguez08}. Clearly, a large
coupling is the key property to be exploited in device applications \cite{Fiebig05}.

It is now expedient to express \equ{motion} in terms of $D$ and $B$. \bea  \left(\begin{array}{c} E \\ H \end{array} \right) &=&  \rr^{-1}(\infty)  \left[  \left(\begin{array}{c} D \\ B \end{array} \right)  - \frac{4 \pi}{\Omega} \sum_n \zz_n u_n \right] \label{inverse} \\  - \omega^2 u_n &=& - \omega_n^2 u_n - \frac{4 \pi}{\Omega} \zz_n^\dagger \rr^{-1}(\infty) \sum_{n'} \zz_{n'} u_{n'}  \nn &+&  \zz_n^\dagger \rr^{-1}(\infty)  \left(\begin{array}{c} D \\ B \end{array} \right). \label{motion2} \eea
As explained above, in the  LO modes  $D$=0 and $B$=0 by definition~\cite{rap_a30}. The first line of \equ{motion2}  clearly shows that the LO eigenmodes are in general different from the TO ones; an explicit $N \times N$ diagonalization is needed  in order to find the LO frequencies. We indicate with $\tilde{\omega}_n$ these frequencies, and we also transform the lattice coupling vectors to the LO eigenmodes as $\rr^{-1}(\infty)  \zz_n\rightarrow \tilde{\zz}_n$, in order to write \equ{motion2} as \[  - \omega^2 u_n = - \tilde{\omega}_n^2 u_n +  \tilde{\zz}_n^\dagger  \left(\begin{array}{c} D \\ B \end{array} \right) . \label{motion3} \] Elimination of the $u_n$'s from \eqs{inverse}{motion3} provides $\rr^{-1}(\omega)$ in the form \bea \rr'^{-1}(\omega) &=& \rr^{-1}(\infty) -  \frac{4 \pi}{\Omega} \sum_n  \frac{\tilde{\zz}_n \tilde{\zz}_n^\dagger}{\tilde{\omega}_n^2  - \omega^2}  \label{ini2} \\ \rr''^{-1}(\omega) &=&  - \frac{4 \pi^2}{\Omega} \sum_n  \tilde{\zz}_n \tilde{\zz}_n^\dagger \, \delta(\tilde{\omega}_n^2 - \omega^2) .  \eea 

Using the above results, we arrive at a very transparent expression for the rhs member of \equ{lst3}: \[  \frac{\evo_{\rm L}}{\evo_{\rm T}} = \frac{\sum_n \frac{1}{\omega_n^2}  \zz_n^\dagger \rr^{-1}(\infty) \zz_n}{ \sum_n \frac{1}{\tilde{\omega}_n^2}  \tilde{\zz}_n^\dagger \rr(\infty) \tilde{\zz}_n } , \label{mean} \] i.e. $\evo_{\rm T}$ is the weighted harmonic mean of the $\omega^2_n$'s, with weights $\zz_n^\dagger \rr^{-1}(\infty) \zz_n$, and $\evo_{\rm L}$ is the weighted harmonic mean of the $\tilde{\omega}^2_n$'s, with weights  $\tilde{\zz}_n^\dagger \rr(\infty) \tilde{\zz}_n$. The two sets of weights are in general different, except when the transverse and longitudinal eigenmodes happen to be the same. However,
``superconvergence'', \equ{iden}, implies the same normalization in any case: \[ \sum_n  \zz_n^\dagger \rr^{-1}(\infty) \zz_n = \sum_n \tilde{\zz}_n^\dagger \rr(\infty) \tilde{\zz}_n .\]
In the special case where a single IR-active mode exists $\evo_{\rm T} = \omega_{\rm T}^2$ and $\evo_{\rm L} = \omega_{\rm L}^2$: this concludes the proof of \equ{rap}.

A well known bound requires the matrix $\rr$ to be positive definite \cite{Fiebig05}.
Since this is based on stability arguments, it only concerns $\rr(0)$ and $\rr(\infty)$: the former is a genuine static property of the real system, while the latter can be regarded as a static property in the infinite nuclear mass limit. At any other frequency the matrix $\rr(\omega)$ accounts for the forced oscillations of the system, which is clearly out of equilbrium. Therefore $\rr(\omega)$ is not required, in general, to be positive definite; because of the same reason,  $\varepsilon(\omega)$ is not a positive real function in ordinary dielectrics \cite{textbook}. 

All of the above results reduce to previously known ones for a magnetically inert material, where the $2 \times 2$ matrix $\rr(\omega)$ has the unique nontrivial entry $\varepsilon(\omega)$. It is worth examining \equ{mean} for $\zeta^*_n = 0$: \[  \frac{\evo_{\rm L}}{\evo_{\rm T}} = \frac{1}{\varepsilon^2(\infty)}\frac{\sum_n \frac{1}{\omega_n^2}  (Z^*_n)^2 }{ \sum_n \frac{1}{\tilde{\omega}_n^2}  (\tilde{Z}^*_n)^2 } . \label{mean2} \] The weights in the harmonic means $(Z^*_n)^2$ and $(\tilde{Z}^*_n)^2$ are the (squared) transverse and longitudinal effective charges, respectively. If (and only if) the transverse and longitudinal eigenmodes  are the same, then $\tilde{Z}^*_n = {Z}^*_n/\varepsilon(\infty)$. Actually, this is the well known relationship between the transverse (a.k.a. Born) end longitudinal (a.k.a. Callen) effective charges.

For an ordinary dielectric the $N$ poles of $\varepsilon(\omega)$ are the TO frequencies $\omega_n$, while the $N$ poles of $1/\varepsilon(\omega)$ are the LO ones  $\tilde{\omega}_n$: see \eqs{ini}{ini2}. 
Since the response is a single-component function,  the poles of $1/\varepsilon(\omega)$ coincide with the zeros of  $\varepsilon(\omega)$. Simple considerations about zeros and poles of  $\varepsilon(\omega)$ eventually lead to the simple expression \[  \frac{\evo_{\rm L}}{\evo_{\rm T}} = \prod_{n=1}^N \frac{\tilde{\omega}_n}{\omega_n} , \] first found in 1961 by Kurosawa~\cite{Kurosawa61,Sievers90}. This result {\it does not} generalize to the ME case, for the good reason that the response is a $2 \times 2$ matrix: the present formulation is based on {\it traces} throughout, and the inverse of the trace bears no simple relationship to the trace of the inverse, at variance with the purely electrical case. 

This Letter addresses the linear relationship between the pairs $(E,H)$ and $(D,B)$ throughout, starting with \equ{resp} onwards. Other pairings are possible.
In particular the choice $(E,B)$ and $(D,H)$ looks like a more natural one for at least two reasons: (i) the microscopic forces are determined by the $(E,B)$ pair \cite{rap_a30}, and (ii)  a relativistic formulation can be elegantly cast in terms of two four-dimensional field tensors whose entries are $(E,B)$ and $(D,H)$, respectively \cite{Hehl08,Hehl09}. Nonetheless, the same choice in the present context would not be a convenient  one. In fact, as explained above (see also Ref. \cite{rap_a30}), in a transverse mode $E=0$ and $B \neq 0$, and in a longitudinal one $D=0$ and $H \neq 0$.

Throughout this Letter we have stressed the formal equivalence of electric and magnetic fields in their coupling to the lattice in MEs. However, the orders of magnitude of electric and magnetic phenomena in condensed matter are not the same. ME effects are notoriously small \cite{Fiebig05}, and the corrections to the LST relationship in most cases are expected to be small as well. In oxides the dielectric constants---either $\varepsilon(\infty)$ or $\varepsilon(0)$---are typically in the range 2 to 10, while $|\mu -1|$ is of the order $10^{-4}$ \cite{nota}; in the most studied linear ME, i.e. Cr$_2$O$_3$, even $\alpha$ is of the order $10^{-4}$ \cite{Iniguez08,Hehl08}. An accurate evaluation of the lhs of \eqs{rap}{lst3} in ``conventional'' ME materials would require a measurement of all the entries of the response matrices $\rr$ to the same absolute error, which could be problematic. More perspicuous effects are expected in nonconventional materials \cite{Fiebig05}, such as those where the ME effect can be tuned \cite{Lee10}.

In conclusion, this Letter shows that the ratio $\varepsilon(0)/\varepsilon(\infty)$ entering the LST relationship must be replaced, for a linear ME, by the lhs of \eqs{rap}{lst3}, whose ingredients are  the full ME responses at frequency $0$ and $\infty$, i.e. static and clamped-ion.  In the most general case the rhs member of our generalized LST relationship is the ratio between spectral moments of the longitudinal and transverse excitations of the medium. It assumes a simple form for a crystalline ME in the harmonic approximation, and finally is identical to the original LST one when only one mode is IR active. The relationship shows that the LO-TO splitting in a ME originates from the coupling of ionic displacements to both electric and magnetic macroscopic fields.

I thank J. \`I\~niguez and D. Vanderbilt for illuminating discussions about magnetoelectrics.  Work supported by the ONR grant  N00014-07-1-1095.


\begin{thebibliography}{10}

\bibitem{Lyddane41}
{ R. H Lyddane, R. G. Sachs, and E. Teller, Phys. Rev. {\bf 59}, 673 (1941)}.

\bibitem{textbook}
{ N. W. Ashcroft and N. D. Mermin, {\it Solid State Physics} (Saunders,
  Philadelphia, 1976), p. 547 onwards}.

\bibitem{Fiebig05}
{ M. Fiebig, J. Phys. D {\bf 38}, R123 (2005)}.

\bibitem{Eerenstein06}
{ W. Eerenstein, N. D. Mathur, and J. F. Scott, Nature (London) {\bf 442}, 759
  (2006)}.

\bibitem{Iniguez08}
{ J. \`I\~niguez, Phys. Rev. Lett. {\bf 101}, 117201 (2008)}.

\bibitem{Hehl08}
{ F. W. Hehl, Y. N. Obukhov, J.-P. Rivera, and H Schmid, Phys. Rev. A {\bf 77},
  022106 (2008)}.

\bibitem{Hehl09}
{ F. W. Hehl, Y. N. Obukhov, J.-P. Rivera, and H Schmid, Eur. Phys. J. {\bf
  71}, 321 (2009)}.

\bibitem{Essin09}
{ A. M. Essin, J. E. Moore, and D. Vanderbilt, Phys. Rev. Lett. {\bf 102},
  146805 (2009)}.

\bibitem{Wojdel09}
{ J. C. Wojdel and J. \`I\~niguez, Phys. Rev. Lett. {\bf 103}, 267205 (2009)}.

\bibitem{Essin10}
{ A. M. Essin, A. M. Turner, J. E. Moore, and D. Vanderbilt, Phys. Rev. B {\bf
  81}, 205104 (2010)}.

\bibitem{Wojdel10}
{ J. C. Wojdel and J. \`I\~niguez, Phys. Rev. Lett. {\bf 105}, 037208 (2010)}.

\bibitem{Kurosawa61}
{ T. Kurosawa, J. Phys. Soc. Jpn. {\bf 16}, 1298 (1961)}.

\bibitem{Cochran62}
{ W. Cochran and R. A. Cowley, J. Phys. Chem. Solids {\bf 23}, 4471 (1962)}.

\bibitem{Lax71}
{ M. Lax and D. F. Nelson, Phys. Rev. B {\bf 4}, 3694 (1971)}.

\bibitem{Gonze97}
{ X. Gonze, Ph. Ghosez, and R. W. Godby, Phys. Rev. Lett. {\bf 78}, 294
  (1997)}.

\bibitem{Barker75}
{ A. S. Barker, Jr. , Phys. Rev. B {\bf 12}, {4071} (1975)}.

\bibitem{Barker75b}
{ A. S. Barker, Jr. and A. J. Sievers, Rev. Mod. Phys. {\bf 47}, S1 (1975)}.

\bibitem{Noh89}
{ T. W. Noh and J. Sievers, Phys. Rev. Lett. {\bf 63},{1800} (1989)}.

\bibitem{Sievers90}
{ A. J. Sievers and J. B. Page, Phys. Rev. B {\bf 41}, 3455 (1990)}.

\bibitem{Altarelli72}
{ M. Altarelli, D. L. Dexter, H. M. Nussenzveig, and D. Y. Smith, Phys. Rev. B
  {\bf 6}, 4502 (1972)}.

\bibitem{units}
{ We adopt Gaussian units, where $E$ and $H$ have the same dimensions, and we
  measure charges in atomic units. In such a way the parameters $Z_n^*$ and
  $\zeta_n^*$ are both dimensionless before being transformed to normal-mode
  coordinates. As for $Z_n^*$, this is standard in the literature}.

\bibitem{rap_a30}
{ R. Resta, J. Phys.: Condens. Matter {\bf 22} 123201 (2010)}.

\bibitem{nota} In the neighborhood of ferromagnetic transitions $|\mu -1|$ may become much larger, while in the presence of soft modes $\varepsilon(0)$ may increase by some orders of magnitude.

\bibitem{Lee10} J. H. Lee {\it et al.}., Nature {\bf 466}, 954 (2010).

\end{thebibliography}

\end{document}